\documentclass[aps,prd,
floatfix,nofootinbib,groupedaddress,showpacs]{revtex4-1}

\usepackage{amsmath}
\usepackage{amssymb}
\usepackage{graphicx}

\usepackage{latexsym}
\usepackage{array}
\usepackage{verbatim}

\usepackage{rotating}
\usepackage{color}

\usepackage[utf8]{inputenc}

\def\lsim{\mathrel{\raise.3ex\hbox{$<$\kern-.75em\lower1ex\hbox{$\sim$}}}}
\def\gsim{\mathrel{\raise.3ex\hbox{$>$\kern-.75em\lower1ex\hbox{$\sim$}}}}

\newcommand{\be}{\begin{equation}}
\newcommand{\ee}{\end{equation}}

\newlength{\absize}
\def\lsim{\mathrel{\rlap{\raise 2.5pt \hbox{$<$}}\lower 2.5pt
\hbox{$\sim$}}}

\newcommand{\Lumint}{{\cal L}_{\rm int}}

\allowdisplaybreaks

\begin{document}

\title{
Improved bounds on $W$-$W'$ mixing with ATLAS resonant $WZ$
production data \\ at the LHC at $\sqrt{s}=13$ TeV }
\author{I.~A. Serenkova}
\email{Inna.Serenkova@cern.ch}
\affiliation{The Abdus Salam ICTP Affiliated
Centre, Technical University of Gomel, 246746 Gomel, Belarus}
\author{P. Osland}
\email{Per.Osland@uib.no}
\affiliation{Department of Physics and
Technology, University of Bergen, Postboks 7803, N-5020 Bergen,
Norway}
\author{A.~A. Pankov}
\email{pankov@ictp.it}
\affiliation{The Abdus Salam ICTP
Affiliated Centre, Technical University of Gomel, 246746 Gomel,
Belarus}
\affiliation{Institute for Nuclear Problems, Belarusian
State University, 220030 Minsk, Belarus}
\affiliation{Joint
Institute  for Nuclear Research, Dubna 141980 Russia}

\date{\today}

\begin{abstract}
New charged vector bosons $W'$ decaying into gauge boson pairs
$WZ$ are predicted in many scenarios of new physics, including
models with an extended gauge sector (EGM).
Due to the large variety of models (other unification
groups, models with Supersymmetry, Little Higgs Models,
Extra Dimensions) the more general EGM approach is here considered.
For what concerns $W'$-production,
these models are parametrised by two parameters, the
$W^\prime$ mass $M_{W^\prime}$ and the $W$-$W^\prime$ mixing parameter
$\xi$. The diboson $WZ$ production allows to place
stringent constraints on this mixing angle and the $W'$
mass, which we determine and present for the first time by using
data from $pp$ collisions at $\sqrt{s}=13$ TeV recorded by the
ATLAS detector at the CERN LHC, with integrated luminosity of
 36.1~fb$^{-1}$. By comparing the experimental limits to
the theoretical predictions for the total cross section of $W'$
resonant production and its subsequent decay into $WZ$  pairs, we
show that the derived constraints on the mixing angle 
are rather severe, between $10^{-4}$ and $10^{-3}$,
i.e., greatly improved with respect to those derived from the
global analysis of electroweak data which yield $\xi\lsim
10^{-2}$. We combine  the limits derived from $WZ$ production data
with those obtained from the $W'\to e\nu$ process in order to
significantly extend the exclusion region in the $M_{W'}$--$\xi$
parameter plane and obtain the most stringent exclusion limits to date.
We present the combined allowed parameter space
for the EGM $W'$ boson after incorporating indirect
constraints from low energy electroweak  data, direct search
constraints from Tevatron and from the LHC Run~I with 7 and 8~TeV
as well as at Run~II with 13~TeV data.
\end{abstract}

\maketitle

\section{Introduction. The $W'$ EGM as a general approach} \label{sec:I}

Many extensions to the Standard Model (SM) predict the existence
of charged and neutral, heavy gauge bosons that could be
discovered at the Large Hadron Collider (LHC). In the simplest models these
particles are considered copies of the SM $W$ and $Z$
bosons and are commonly referred to as $W'$ and $Z'$ bosons
\cite{Tanabashi:2018oca}. The Sequential Standard Model (SSM)
\cite{Altarelli:1989ff} posits a $W'_{\rm SSM }$ boson with
couplings to fermions that are identical to those of the SM $W$
boson but for which the coupling to $WZ$ is absent.
The SSM has been used as a reference point for experimental $W'$
boson searches for decades, the results can be re-interpreted in the context of
other models of new physics, and it is useful for comparing the
sensitivity of different experiments.

At the LHC, a promising way to search for heavy $W^\prime$
bosons is through their single production as an $s$-channel
resonance with their subsequent leptonic decays
\begin{equation}
pp\to W^\prime X \to \ell\nu X, \label{proclept}
\end{equation}
where in what follows, $\ell=e,\mu$ unless otherwise stated. The Feynman diagram for
the $W'$ boson production and its dilepton decay at the parton
level is illustrated in  Fig.~\ref{figln}.
This process (\ref{figln}) offers the simplest event topology for the
discovery of a $W^\prime$ with a large production rate and a clean
experimental signature. These channels are among the most
promising early discoveries at the
LHC~\cite{Aaboud:2017efa,Sirunyan:2018mpc}. There have also been many
theoretical studies of $W^\prime$
boson searches \cite{Schmaltz:2010xr,Grojean:2011vu,Boos:2011ib,Jezo:2012rm,Cao:2012ng,
Boos:2013cxa,Bakhet:2014ypa,Dobrescu:2015yba,Dev:2015pga,Baskakov:2018adc}
at the LHC.

\begin{figure}[!htb]
\refstepcounter{figure} \label{figln}
 \addtocounter{figure}{-1}
\begin{center}
\includegraphics[scale=0.80]{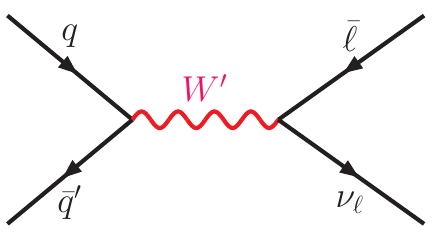}
\end{center}
\vspace*{0.cm} \caption{Leading-order (LO) Feynman diagram of
$W'$ boson production and dilepton decay. The final state
shown denotes both the ($\bar{\ell}\nu_{\ell}$) state and its
charge conjugated one.}
\end{figure}

The ATLAS  and CMS collaborations set limits on the $W'$
production cross section times branching fraction in the process
(\ref{proclept}), $\sigma(pp\to W'X)\times \text{BR}(W'\to \ell\nu)$, for
$M_{W'}$ in the 0.15 TeV -- 6 TeV range. The most stringent
limits on the mass of a $W'_{\rm SSM}$ boson to date come from the
searches in the $W'\to e\nu $ and $W'\to \mu\nu$ channels by the
ATLAS and CMS collaborations using data taken at $\sqrt{s} = 13$
TeV. The ATLAS and CMS analyses were based on data corresponding to
an integrated luminosity of 36.1~fb$^{-1}$ and 35.9~fb$^{-1}$ and
set a 95\% confidence level (CL) lower limit on the $W'_{\rm SSM}$
mass of 5.2~TeV \cite{Aaboud:2017efa,Sirunyan:2018mpc}.

An alternative $W'$ search channel is the diboson one
\begin{equation}
pp\to W^\prime X\to WZ X. \label{procWZ}
\end{equation}
The study of gauge boson pair production offers a powerful test of the
spontaneously broken gauge symmetry of the SM and can be used as a
probe for new phenomena beyond the SM.

Massive resonances that can decay to gauge boson pairs  are
predicted by many scenarios of new physics, including extended
gauge models (EGM) \cite{Altarelli:1989ff,Eichten:1984eu}, models
of warped extra
dimensions~\cite{Randall:1999ee,Davoudiasl:2000wi}, technicolour
models~\cite{Lane:2002sm,Eichten:2007sx} associated with the existence
of technirho and other technimesons, and more generic composite
Higgs models \cite{Agashe:2004rs, Giudice:2007fh,Liu:2018hum}, etc.
 Searches for exotic heavy particles that decay into $WZ$ pairs are complementary
to searches in the leptonic channels $\ell\nu$ of process
(\ref{proclept}). Moreover, there are models in which the
$W'$ couplings to SM fermions are suppressed, giving rise to a
fermiophobic $W'$ with an enhanced coupling to $W$ and $Z$
bosons~\cite{Tanabashi:2018oca, He:2007ge}. It is therefore important to
search for $W'$ bosons also in the $WZ$ final state.

Given the large variety of models which predict new heavy charged
gauge bosons, it is a natural approach to use a simplified ansatz
for such a search \cite{DeGuio:2012ixw}. After a discovery of
signatures associated to a new boson, detailed studies can be
carried out to distinguish between these models and to determine
whether the boson belongs to one of the theoretically motivated
models such as an EGM, models of warped extra dimensions,
technicolour, composite Higgs models, a Little Higgs model, a
Left-Right Symmetric model or a totally different one. Following
the traditions of direct searches at hadron colliders, such
studies are based on the model first proposed in
Ref.~\cite{Altarelli:1989ff}.

As mentioned above, in the SSM, the coupling constants of the
$W'$ boson with SM fermions are the direct transcription of the
corresponding SM couplings, while the $W'$ coupling to $WZ$ is
absent, $g_{W'WZ}=0$ \cite{Aad:2014xka,Aad:2015ipg}.
Note that this suppression may arise in an EGM in a natural
manner: if the new gauge bosons and the SM ones belong to
different gauge groups, a vertex such as $W'WZ$ is forbidden. They
can only occur after symmetry breaking due to mixing of the gauge
eigenstates. Triple gauge boson couplings (such as $W'WZ$) as well
as the vector-vector-scalar couplings (like $W'WH$) arise from the
symmetry breaking and may contribute to the $W'$ decay. The
vertices are then suppressed by a factor of the order of
$(M_W/M_{W'})^2$.

Heavy spin-1 resonances are a generic prediction
of many EGMs. In particular,  the EGM introduces extra, heavy, charged
$W'$ and neutral $Z'$
bosons with SM couplings to fermions and with modified $W$-$W'$
mixing-induced coupling of the heavy $W'$ to $WZ$. Here, we
concentrate on a study of these mixing effects
in the decay of $W'$ to $WZ$.\footnote{ Analogous analysis of the
$Z$-$Z'$ mixing proposed in \cite{Andreev:2014fwa} and recently
performed on the basis of resonant diboson production data in
$pp\to Z'X\to W^+W^-X$ at the ATLAS and CMS can be found in e.g.
\cite{Osland:2017ema,Bobovnikov:2018fwt}.}

In an EGM, the trilinear gauge boson coupling is modified by a mixing
factor
\begin{equation} \label{Eq:define-xi}
\xi={\cal C} \times (M_W/M_{W'})^2,
\end{equation}
where ${\cal C}$ is a scaling constant that sets the coupling
strength. Note that the EGM can be parametrized either in terms of
$(M_{W'},{\cal C})$ or in terms of $(M_{W'},\xi)$. Specifically,
in an EGM the standard-model trilinear gauge boson coupling
strength $g_{WWZ}$ ($=e\cot\theta_W$), is replaced by
$g_{W'WZ}=\xi\cdot g_{WWZ}$. Following the parametrization of the
trilinear gauge boson coupling $W'WZ$ presented in
\cite{Aaltonen:2010ws} for the analysis and interpretation of the
CDF data on $p\bar{p}\to W'X\to WZX$, expressed in terms of
two free parameters,\footnote{Such a $W'$, described in terms of
two parameters is here referred to as the EGM boson.} $\xi$ and
$M_{W'}$, we will set, for the first time, $W'$ limits as
functions of the mass $M_{W'}$ and mixing factor $\xi$ by using
the ATLAS resonant diboson production data \cite{Aaboud:2017fgj}
collected at a center of mass energy of $\sqrt{s}=13$ TeV. The
presented analysis in the EGM with two free parameters is more
general than the previous ones where the only parameter is the
$W'$ mass. As to the SSM, we have  $W'_{\rm SSM}\equiv W'_{\rm
EGM}({\xi}=0)$.

Previous searches for an EGM $W'$ resonance in the $WZ$ channel have
been carried out using $p\bar{p}$ collision data at
$\sqrt{s}=1.96$ TeV at the Tevatron and $pp$ collision data at
$\sqrt{s}=$7 and 8~TeV at the LHC. Early results from the
Tevatron~\cite{Aaltonen:2010ws, Abazov:2009eu} derived by the CDF  and
D0 collaborations have put limits, at the 95\% CL, on the mass of
a $W'$ boson between 285 and 516 GeV and between 188
and 520 GeV, respectively. The ATLAS and CMS collaborations
have also set exclusion bounds on the production and decay of a
$W'$ boson, in searches using the $\ell\nu\, \ell'\ell'$  channel,
the ATLAS ~\cite{Aad:2014pha} and CMS~\cite{Khachatryan:2014xja}
collaborations have excluded, at the 95\% CL, EGM (${\cal C}=1$)
$W'$  bosons decaying to $WZ$ for $W'$ masses below 1.52  TeV and
1.55 TeV, respectively. Here $\ell'$ stands for an electron or
muon.  In addition the ATLAS Collaboration has excluded EGM
(${\cal C}=1$) $W'$ bosons for masses below 1.59 TeV using the
$\ell\ell\, q{q}$~\cite{Aad:2014xka} channel, and below 1.49 TeV
using the $\ell\nu\,q{q}$~\cite{Aad:2015ufa} channel. These have
also been excluded with masses between 1.3 and 1.5 TeV and below
1.7 TeV by the ATLAS~\cite{Aad:2015owa} and
CMS~\cite{Khachatryan:2014hpa} collaborations, respectively, using
the fully hadronic final state. To improve the sensitivity to new
diboson resonances  in the context of the EGM (${\cal C}=1$) in order
to set the strongest exclusion bounds on the $W'$ masses, the  fully
leptonic, semi-leptonic and fully hadronic channels at 8 TeV  were
combined \cite{Aad:2015ipg}. The result of this combination was
interpreted using the EGM $W'$ model with ${\cal C}=1$  as a
benchmark. The observed lower limit on the  $W'$ mass was found to
be 1.8 TeV. The various decay channels generally differ in
sensitivity in different mass regions. The fully leptonic channel,
in spite of a lower branching ratio, is expected to be
particularly sensitive to low-mass resonances as it has lower
backgrounds.

The strongest lower limit on the $W'$ mass  set at 13 TeV is
$M_{W'}>3.2$ TeV \cite{Tanabashi:2018oca} in the context of the heavy
vector-triplet (HVT) model of
``weekly-coupled scenario A'' \cite{Pappadopulo:2014qza} . The HVT generalises a large number of
models that predict spin-1 charged ($W'$) and neutral ($Z'$)
resonances. Such models can be described in terms of just a few
parameters: two coefficients $c_\mathrm{F}$ and $c_\mathrm{H}$,
scaling the couplings to fermions, and to the Higgs and
longitudinally polarized SM vector bosons respectively, and the
strength $g_{\rm V}$ of the new vector boson interaction. Two
benchmark models are considered in the HVT scenario. We are
interested in one of them, referred to as HVT model-A, with
$g_{\rm V} = 1$ because of its similarity to the EGM (${\cal
C}=1$) $W'$ model, that have comparable branching fractions to
fermions and gauge bosons.

Limits were also set on the EGM~$W'$~boson coupling strength
scaling factor ${\cal C}$, as functions of $M_{W'}$, within the EGM framework
\cite{Aaltonen:2010ws,Abazov:2009eu,Aad:2013pdy,Chatrchyan:2012kk,Khachatryan:2014xja}.
It was shown that if the coupling between the $W'$ boson and
$WZ$ happens to be stronger (weaker) than that predicted by the
EGM with ${\cal C}=1$, the observed and expected limits will be
more stringent (relaxed).

The properties of $W'$ bosons are also constrained by measurements
of EW processes at low energies, i.e., at energies much below the mass
$M_{W'}$. Such bounds on the $W$-$W'$ mixing are mostly due to the
change in $W$ properties compared to the SM predictions. These
measurements show that the mixing angle $\xi$ between the gauge
eigenstates must be smaller than about 10$^{-2}$ \cite{Tanabashi:2018oca}.

In this work, we derive bounds on a possible new charged spin-1
resonance ($W^\prime$) in the EGM framework from the available ATLAS data
on $WZ$ pair production \cite{Aaboud:2017fgj}.  The search was conducted for a new $W'$
resonance decaying into a $WZ$ boson pair, where the $W$ boson
decays leptonically ($W\to\ell\nu$ with $\ell=e,\mu$) and the $Z$
boson decays hadronically ($Z\to q\bar{q}$ with $q$ quarks). We
present results as constraints on the relevant $W$-$W^{\prime}$
mixing angle $\xi$ and on the mass $M_{W'}$ and display the
combined allowed parameter space for the EGM $W'$ boson,
showing also indirect constraints from electroweak precision data
(EW), direct search constraints from the Tevatron and from the LHC with 7
and 8 TeV as well as with 13 TeV data.

The paper is organized as follows.  In
Sec.~\ref{sect-cross} we summarize the relevant cross section and
study the $W'\to WZ$ width in the EGM. Then, in
Sec.~\ref{sect-constraints} we show the resulting constraints on
the $M_{W'}$-$\xi$ parameter space obtained from diboson and
dilepton processes.  In Sec.~\ref{sect:overall_constraints} we
collect and compare the indirect constraints obtained from
electroweak precision data, direct search constraints derived from
the diboson process at the Tevatron and at the LHC.
Also, we explore  the role of the dilepton
process in reducing the excluded area in the $W'$ parameter plane,
and in Sec.~\ref{sect:conclusions} we conclude.

\begin{figure}[!htb]
\refstepcounter{figure} \label{figWZ} \addtocounter{figure}{-1}
\begin{center}
\vspace*{0.cm}
\includegraphics[scale=0.80]{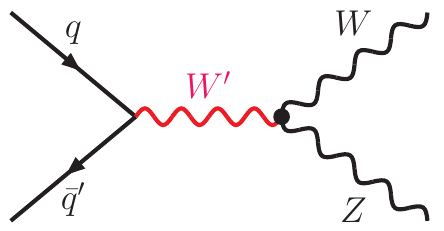}
\end{center}
\vspace*{0.cm} \caption{Lowest-order Feynman diagram for the $W'$ boson
production and decay to the diboson $WZ$ final state.}
\end{figure}

\section{$W'$ production and decay at the LHC }
\label{sect-cross}
 At lowest order within the EGM, $W'$ production and decay into $WZ$ in
proton-proton collisions occurs through quark-antiquark
interactions in the $s$-channel, as illustrated by the Feynman
diagram shown in Fig.~\ref{figWZ}.\footnote{This may however be considered an {\it effective} diagram, in the sense that an underlying theory may generate the $W^\prime WZ$ vertex at loop order.}
The  cross section of process (\ref{procWZ})  at the LHC can be
observed through resonant pair production of gauge bosons $WZ$.
Using the narrow width approximation (NWA), one can factorize the
process (\ref{procWZ}) into the $W'$ production and the $W'$
decay,
\begin{equation}
\sigma(pp\to W' X\to WZX)  = \sigma(pp\to W'X) \times \text{BR}(W' \to
WZ)\;. \label{TotCr}
\end{equation}
Here, $\sigma(pp\to W' X)$ is the  total (theoretical) $W'$
production cross section and
$\text{BR}(W' \to WZ)=\Gamma_{W'}^{WZ}/\Gamma_{W'}$ with
$\Gamma_{W'}$ the total width of $W'$. The cross section
$\sigma(pp\to W' X)$ for the inclusive $W'$ production $pp\to W'X$
is derived from
the quark subprocess $q\bar{q}'\to W'$ which can be written
as \cite{Barger:1987nn}:
\begin{equation}
\hat\sigma(q\bar{q}'\to W')=\frac{\pi\vert V_{qq'}\vert^2}{4}\,
g^2\delta(\hat{s}-M_{W'}^2). \label{hat}
\end{equation}
Here, the weak coupling constant $g=e/\sin\theta_W$, and $V_{qq'}$ is the
Cabibbo-Kobayashi-Maskawa (CKM) matrix element  connecting quark
$q$ and antiquark $\bar{q}'$. The hadronic cross section can be
obtained by the summation over all contributing quark-antiquark
combinations and integration over the momentum fractions
\cite{Altarelli:1989ff,Nuss:1996dg},
\begin{equation}
\sigma(p_1p_2\to W' X) = \frac{K}{3}\int_0^1 dx_1 \int_0^1 dx_2
\sum_{q,\bar q'} [f_{q|p_1}(x_1, M_{W'}^2)f_{\bar
{q}^\prime|p_2}(x_2,M_{W'}^2)]\,\hat\sigma(q\bar{q}'\to W').
\label{cross}
\end{equation}
With $\hat s$ the parton subprocess c.m.\ energy squared,
and $s$ the proton-proton c.m.\ energy squared,
it is assumed that $\hat{s}=x_1\,x_2s=M_{W'}^2$ is the
appropriate scale of the quark distributions.
The coefficient of
$1/3$ in front of Eq.~(\ref{cross}) is a color factor.
Furthermore, $f_{q|p_1}(x_{1},M_{W'}^2)$ and $f_{\bar
q'|p_2}(x_2,M_{W'}^2)$ are quark and antiquark momentum distribution
functions for the two protons, with $x_{1,2}$ the parton fractional
momenta, related to the rapidity $y$ via
$x_{1,2}=(M_{W'}/\sqrt{s})\exp(\pm y)$.
The $K$ factor accounts for higher-order QCD
contributions. For the numerical computation, we use the CTEQ-6L1
parton distributions \cite{Pumplin:2002vw} with the factorization
and renormalization scales $\mu_{\rm F}^2=\mu_{\rm
R}^2=M_{W'}^2=\hat{s}$. The obtained constraints presented in the
following are numerically not significantly modified when
$\mu_{\rm F,R}$ is varied in the range from $\mu_{\rm F,R}/2$ to
$2\mu_{\rm F,R}$.

\begin{figure}[htb]
\refstepcounter{figure} \label{fig3} \addtocounter{figure}{-1}
\begin{center}
\includegraphics[scale=0.72]{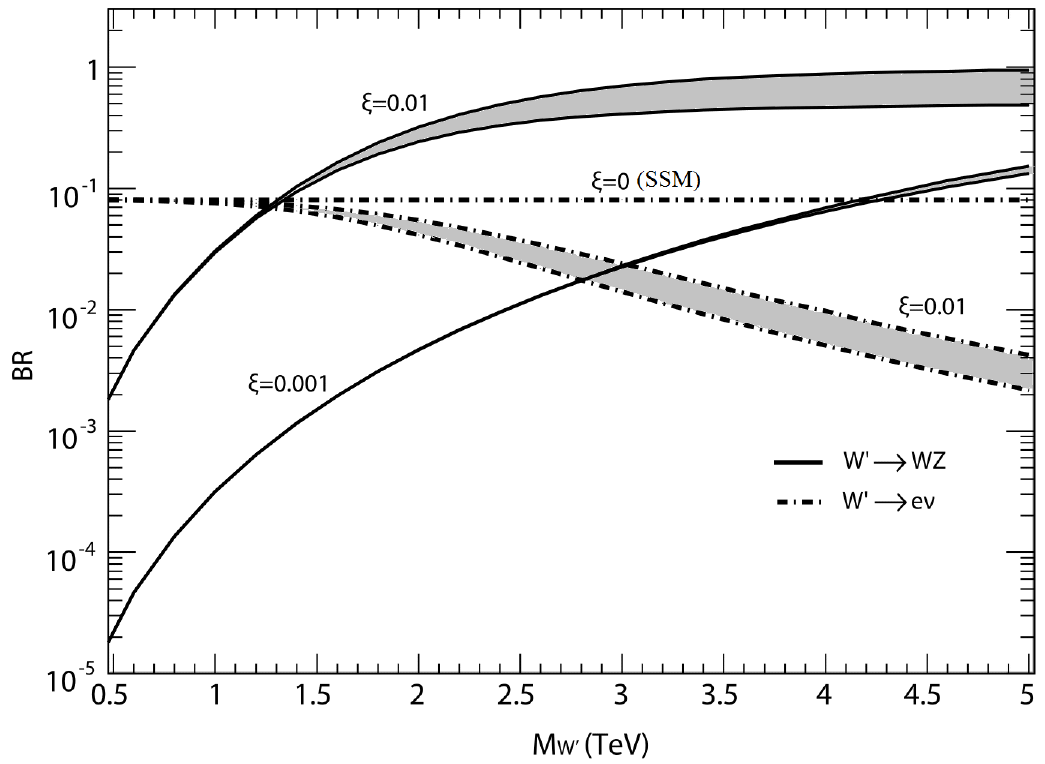}
\end{center}
\caption{Branching fraction $\text{BR}(W'\to WZ)$
(solid) vs $M_{W'}$ in the
EGM for non-zero $W$-$W'$ mixing
factor $\xi=10^{-3}$ and $10^{-2}$.
For the $W'\to e\nu$ mode (dot-dashed), $\text{BR}(W'\to e\nu)$ for
$\xi=0$ ($W'_{\rm SSM}$)  and $\xi=0.01$ ($W'_{\rm EGM}$) are shown.
The shaded bands indicate the uncertainty resulting from
the inclusion of the $WH$ decay mode, the upper and lower bounds correspond to
the assumptions $\Gamma_{W'}^{WH}=0$ and $\Gamma_{W'}^{WH}=\Gamma_{W'}^{WZ}$, respectively.}
\end{figure}

In the EGM the $W'$ bosons can decay into the SM fermions, gauge
bosons ($WZ$), or a pair of an SM boson and a Higgs boson. In the calculation
of the total width $\Gamma_{W'}$ we consider the following
channels: $W'\to f{\bar{f}}^\prime$, $WZ$, and $WH$, where $H$ is
the SM Higgs boson and $f$ are the SM fermions ($f=\ell,\nu,q$).
In this study only left-handed neutrinos are considered while
possible right-handed neutrinos are assumed to be kinematically
unavailable as final states. Also, throughout the paper we shall
ignore the couplings of the $W'$ to beyond-SM particles such as
SUSY partners and any exotic fermions in the theory, which
may increase the width of the $W'$ and hence lower the
branching ratio into a $WZ$ pair. As a result, the total decay
width of the $W^\prime$ boson is taken to be
\begin{equation}\label{gamma}
\Gamma_{W'} = \sum_f \Gamma_{W'}^{f\bar{f}'} + \Gamma_{W'}^{WZ} +
\Gamma_{W'}^{WH}.
\end{equation}
The presence of the two last decay channels,  which are often
neglected at low and moderate values of $M_{W'}$, is due to
$W$-$W'$ mixing which is constrained to be tiny.
In particular, for the
range of $M_{W'}$ values below $\sim 1.0-1.5$ TeV,  the dependence
of $\Gamma_{W'}$ on the values of $\xi$ (within its allowed range)
induced by $\Gamma_{W'}^{WZ}$ and $\Gamma_{W'}^{WH}$ is
unimportant because $\sum_f \Gamma_{W'}^{f\bar {f'}}$ dominates
over diboson partial widths. Therefore, in this mass range, one
can approximate the total width as $\Gamma_{W'} \approx \sum_f
\Gamma_{W'}^{f\bar {f'}}$,
 where the sum runs over SM fermions only.

 Due to the assumption
 that the $W'$ and the $W$ have identical couplings to the SM fermions,
 the total $W'$ width can be  expressed in terms of the $W$ width $\Gamma_W$.
For $W'$ masses below  $m_t+m_b\simeq 180$ GeV the kinematically
allowed decay channels are identical for the SM $W$ and the total
width reads as $\Gamma_{W'}=(M_{W'}/M_W)\,\Gamma_W$. For $W'$
masses above 180 GeV the decay $W'\to t b$ opens. Since the phase
space is enlarged it results in an increase of the $W'$ width by a
factor of 4/3, namely $\Gamma_{W'}=(4M_{W'}/3M_W)\,\Gamma_W$. In
this $W'$ mass range limited by the upper bound of about
1.5~TeV, a predicted branching fraction $\text{BR}(W'\to
l\nu)=1/12$ (or about 8.2\%) for each of the leptonic channels
studied, as illustrated in Fig.~\ref{fig3} for the case of
vanishing $W$-$W'$ mixing ($\xi=0$). Under these assumptions
(for decays to massless SM fermions),
the total width scales with the mass as
$\Gamma_{W'}=(g^2/4\pi)M_{W'}$ \cite{Barger:1987nn}.
Allowing also for final-state QED and QCD corrections,
one arrives at
$\Gamma_{W'}\approx 3.5\% \times M_{W'}$ (see, e.g. \cite{ATLAS:2014wra}).

For heavier $W'$ bosons, the diboson decay channels, $WZ$ and
$WH$,  start to play an important role because of their
significant contribution to  the $W'$ total decay width,
$\Gamma_{W'}$, and to the branching ratio $\text{BR}(W' \to WZ)$,
we are no longer able to ignore them.
To be specific, we take an approach as model-independent as
possible, and  for numerical illustration show our results in two
simple scenarios. In the first scenario,
we treat the model as effectively having a suppressed partial
width of $W'\to WH$ with respect to that of $W'\to WZ$, i.e.
$\Gamma_{W'}^{WH}\ll\Gamma_{W'}^{WZ}$, so that one can ignore the
former, taking $\Gamma_{W'}^{WH}\simeq 0$. In this case, numerical
results with our treatment will serve as an upper bound on the
size of the signal. The second scenario assumes that
both partial widths are comparable, $\Gamma_{W'}^{WH}\simeq
\Gamma_{W'}^{WZ}$ for heavy $M_{W'}$ as required by the
equivalence theorem \cite{Chanowitz:1985hj}. In particular, the
equivalence theorem requires that $W'$ decays into fields that are
part of the same Higgs doublet (e.g., longitudinal $Z$ and $H$)
have equal decay widths up to electroweak symmetry breaking
effects and phase-space factors \cite{Dobrescu:2015yba}.
While the equivalence theorem
 might suggest a value for $\text{BR}(W'\to
WH)$ comparable to $\text{BR}(W'\to WZ)$, the $W'WH$ coupling is
actually quite model dependent \cite{Dobrescu:2015yba}. In the numerical
analysis presented below, we will consider both
scenarios. In the second scennario, $\Gamma_{W'}$ would
be larger, with a suppression of the
branching ratio to $WZ$, and the bounds from LHC (and the ability
for observing the $W$-$W'$ mixing effect) would be reduced.

Note also that for all $M_{W'}$ values of interest for LHC the
width of the $W'$ boson is considerably smaller than the
experimental mass resolution $\Delta M$ ($M\equiv\sqrt{\hat s}$)
for which we adopt the parametrization in reconstructing the
diboson invariant mass of the $WZ$ system, $\Delta M/M\approx 5\%$,
as proposed, e.g., in \cite{Aaboud:2016okv,Sirunyan:2017nrt}.
This condition validates the NWA adopted in this work.

The expression for the partial width of the $W'\to WZ$ decay
channel in the EGM can be written as \cite{Altarelli:1989ff}:
\begin{eqnarray}
\label{GammaWZ}
 \Gamma_{W'}^{WZ}&=&\frac{\alpha_{\rm
em}}{48}\cot^2\theta_W\, M_{W'}
\frac{M_{W'}^4}{M_W^2M_Z^2}\left[\left(1-\frac{M_Z^2-M_W^2}{M_{W'}^2}\right)^2
-4\,\frac{M_W^2}{M_{W'}^2}\right]^{3/2} \\ \nonumber
&& \times\left[
1+10 \left(\frac{M_W^2+M_Z^2}{M_{W'}^2}\right) +
\frac{M_W^4+M_Z^4+10M_W^2M_Z^2}{M_{W'}^4}\right]\cdot\xi^2.
\end{eqnarray}
As one can see from Eqs.~(\ref{gamma}) and   (\ref{GammaWZ}), in
the first scenario where $\Gamma_{W'}^{WH}=0$, for a fixed mixing
factor $\xi$ and at large $M_{W'}$, where $\Gamma_{W'}^{WZ}$
dominates over $\sum_f \Gamma_{W'}^{f\bar {f'}}$  the total width
increases rapidly with the $W'$ mass because of the quintic
dependence on the $M_{W'}$ mass of the $WZ$ mode,
$\Gamma_{W'}^{WZ}\propto M_{W'}\cdot {M_{W'}^4}/({M_W^2M_Z^2})$,
 which corresponds to the production of
longitudinally polarized $W$ and $Z$ in the decay channel $W'\to
W_LZ_L$. In this case, the $WZ$ mode becomes dominant and
$\text{BR}(W' \to WZ)\to 1$, while the fermionic decay channels
($\sum_f \Gamma_{W'}^{f\bar {f'}}\propto M_{W'}$) are increasingly
suppressed. However, in the second scenario with
$\Gamma_{W'}^{WH}=\Gamma_{W'}^{WZ}$, $\text{BR}(W' \to WZ)\to 0.5$
when $M_{W'}$ increases, as is demonstrated in
Fig.~\ref{fig3}, in particular for larger allowed value of the mixing
factor, $\xi=0.01$. Also, Fig.~\ref{fig3} shows that the branching
ratio of $W'$ to fermions, e.g. $\text{BR}(W'\to e\nu)$, decreases
as $\xi$ increases. This is opposite to the diboson decay mode of
$W'\to WZ$ where the branching ratio increases as $\xi$ increases.

\section{Constraints on $W'$ from the diboson and dilepton
processes } \label{sect-constraints}

\subsection{$W$-$W'$ mixing effects in  $W'\to WZ$}
\label{subsect-diboson}

Here, we present an analysis, employing the most recent measurements of
diboson processes provided by ATLAS \cite{Aaboud:2017fgj}.  In
Fig.~\ref{fig4}, we show the observed and expected $95\%$ C.L.
upper limits on the production cross section times the branching
fraction, $\sigma_{95\%}\times \text{BR}(W'\to WZ)$, as a function
of the $W'$ mass, $M_{W'}$.
The expected upper limit set on the signal cross section is
the greatest value of the signal cross section that is not excluded
with 95\% confidence.
 The data analyzed comprises $pp$
collisions at $\sqrt{s}=13$ TeV, recorded by the ATLAS (36.1
fb$^{-1}$) detector \cite{Aaboud:2017fgj} at the LHC. As mentioned
above, ATLAS analyzed the $WZ$ production in the process
(\ref{procWZ}) through the semileptonic final states.

\begin{figure}[htb!]
\refstepcounter{figure} \label{fig4} \addtocounter{figure}{-1}
\begin{center}
\includegraphics[scale=0.75]{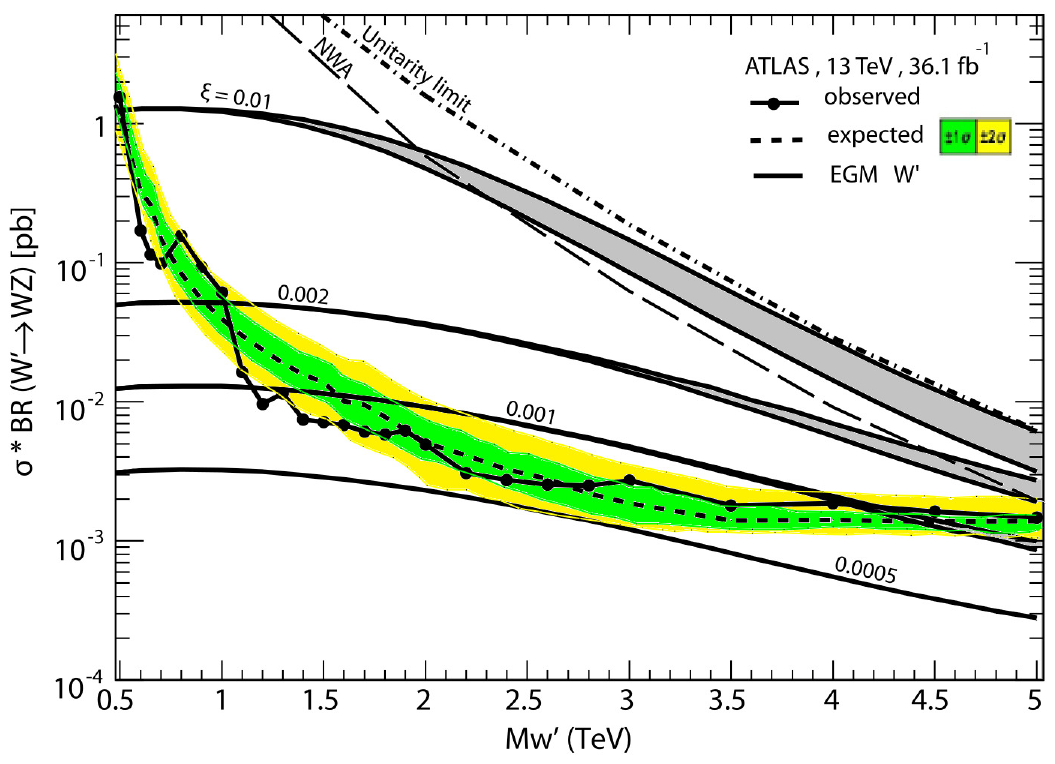}
\end{center}
 \caption{Observed and expected $95\%$ C.L. upper
limits on the production cross section times the branching
fraction, $\sigma_{95\%}\times \text{BR}(W'\to WZ)$,
as a function of the $W'$ mass, $M_{W'}$,
showing ATLAS data for $36.1~\text{fb}^{-1}$ \cite{Aaboud:2017fgj}.
Theoretical production cross sections
$\sigma(pp\to W'X)\times \text{BR}(W'\to WZ)$ for the ${\rm EGM}$ are
calculated from PYTHIA~8.2 with a $W'$ boson mass-dependent
$K$-factor used to correct for NNLO QCD cross sections,
and given by solid curves, for mixing factor $\xi$ ranging from
0.01 and down to 0.0005. The shaded bands are defined like in Fig.~\ref{fig3}.
The area lying below the
long-dashed curve labelled by NWA corresponds to the region where
the $W'$ resonance width is predicted to be less than 5\% of the
resonance mass, in which the narrow-resonance assumption is
satisfied. The lower boundary of the region excluded by the
unitarity violation arguments is also indicated by the dot-dashed curve.}
\end{figure}

Then, for $W'$ we compute the LHC theoretical production cross
section multiplied by the branching ratio into  $WZ$ bosons,
$\sigma (pp\to W'X)\times {\rm BR}(W'\to WZ)$, as a function of the
two parameters ($M_{W'}$, $\xi$), and compare it with the limits
established by the ATLAS experiment, $\sigma_{95\%} \times {\rm
BR}(W'\to WZ)$. Our strategy in the present analysis is to adopt
the SM backgrounds that have been carefully evaluated by the
experimental collaboration and simulate only the $W'$ signal. We
set cross section limits on $W'$ as functions of $M_{W'}$
and $\xi$. Our results extend the sensitivity beyond the corresponding
CDF Tevatron  results \cite{Aaltonen:2010ws,
Abazov:2009eu} as well as the ATLAS and CMS sensitivity attained
at 7 and 8 TeV in
\cite{Aaltonen:2010ws,Abazov:2009eu,Aad:2013pdy,Chatrchyan:2012kk,Khachatryan:2014xja}.
Also, for the first time, we set $W'_{\rm EGM}$ limits as functions
of mass $M_{W'}$ and mixing factor $\xi$ at the LHC from the 13 TeV data with
a luminosity of 36.1~fb$^{-1}$.

In Fig.~\ref{fig4}, the inner (green) and outer (yellow) bands
around the expected limits represent $\pm 1\sigma$ and $\pm
2\sigma$ uncertainties, respectively. The simulation of signals
for the EGM $W'$ is based on an adapted version of
the leading order (LO) PYHTHIA 8.2
event generator \cite{Sjostrand:2014zea}.
A mass-dependent $K$ factor is used to rescale the LO
PYTHIA prediction to the next-to-next-to-leading-order (NNLO) in
$\alpha_s$. The theoretical $W'$ production cross section $\sigma(pp\to
W'X)$ is scaled to a NNLO calculation in
$\alpha_s$ by ZWPROD \cite{Aad:2014pha, Aad:2015ufa,Hamberg:1990np}, given by solid curves, and shown for
mixing factor $\xi$ ranging from 0.01 and down to 0.0005. The
factorization and renormalization scales are set to the $W'$
resonance mass.

As was explained in connection with Fig.~\ref{fig3}, the upper (lower) boundary
of the  shaded  areas correspond to a
scenario where the contribution of the decay
channel $W'\to WH$ to the total $W'$ decay width of
Eq.~(\ref{gamma}) is $\Gamma_{W'}^{WH}=0$
($\Gamma_{W'}^{WH}=\Gamma_{W'}^{WZ}$). The area below the
long-dashed curve labelled ``NWA'' corresponds to the region where
the $W'$ resonance width is predicted to be less than 5\% of its
mass, corresponding to the best detector resolution of
the searches, where the narrow-resonance assumption is satisfied.
In addition, in Fig.~\ref{fig4} we plot a curve labelled
``Unitarity limit'' that corresponds to the unitarity bound (see,
e.g. \cite{Alves:2009aa} and references therein, where it was
shown that the saturation of unitarity in the elastic scattering
$W^\pm Z\to W^\pm Z$ leads to the constraint ${g_{W'WZ}}_
\text{max}=g_{WWZ}\cdot M_Z^2/(\sqrt{3}\,M_{W'}\,M_W)$).
This bound was
obtained under the assumption that the couplings of the $W'$ to
quarks and to gauge bosons have the same Lorentz structure as
those of the SM, but with a rescaled strength.

\begin{figure}[!t]
\refstepcounter{figure} \label{fig5} \addtocounter{figure}{-1}
\begin{center}
\includegraphics[scale=0.7]{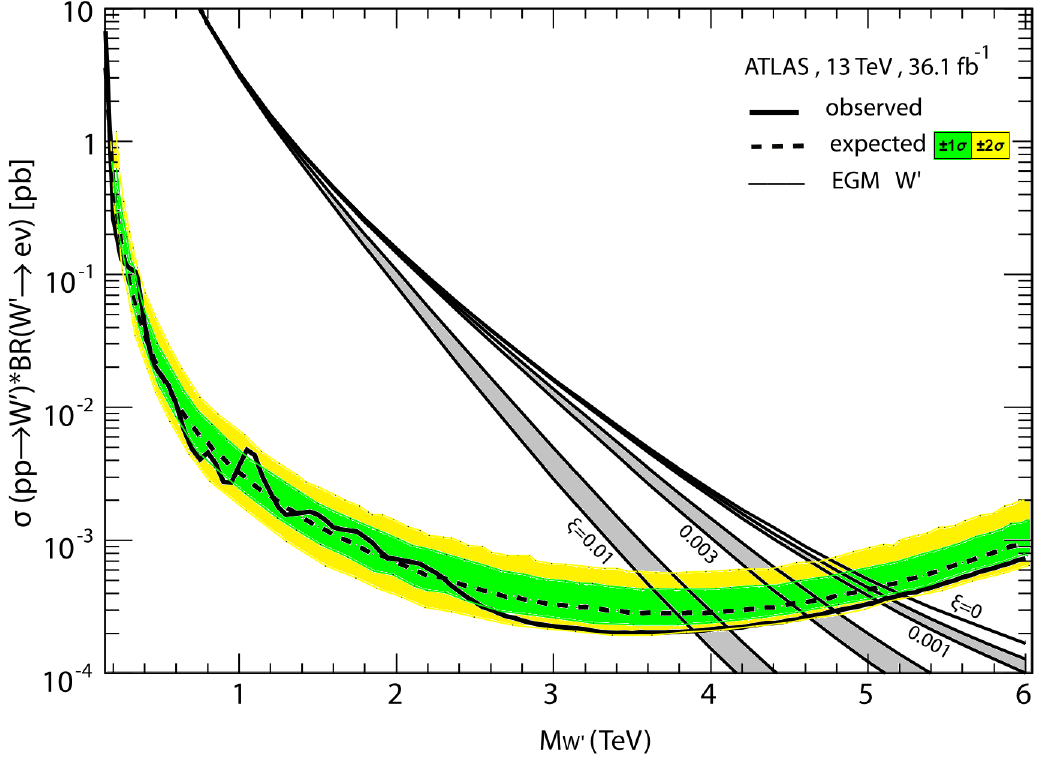}
\end{center}
\caption{Solid (dashed): observed (expected) $95\%$
C.L. upper bound on the $W'$ production cross section times
branching ratio to two leptons, $\sigma_{95\%}\times
\text{BR}(W'\to e\nu)$, obtained at the LHC with integrated
luminosity $\Lumint$=36.1\, fb$^{-1}$ by the ATLAS collaboration
\cite{Aaboud:2017efa}. Thin lines: theoretical production cross
section $\sigma(pp\to W'X) \times {\rm BR}(W'\to e\nu)$
 for the EGM $W'$ boson, calculated from PYHTHIA~8.2 with an NNLO $K$ factor.
These curves in descending order correspond to representative
values of the $W$-$W'$ mixing factor $\xi$ from 0 to 0.01.
The shaded bands are defined like in Fig.~\ref{fig3}.}
\end{figure}

The theoretical curves for the cross sections $\sigma(pp\to W'X)
\times {\rm BR}(W'\to WZ)$, in descending order, correspond to
values of the $W$-$W'$ mixing factor $\xi$ from 0.01 to 0.0005.
The intersection points of the expected (and measured) upper
limits on the production cross section with these theoretical
cross sections for various $\xi$ give the corresponding lower
bounds on ($M_{W'}$, $\xi$), to be displayed  below, in
Sec.~\ref{sect:overall_constraints}.

\subsection{$W$-$W'$ mixing effects in $W'\to \ell\nu$}
\label{subsect-dilept}

The above analysis was for the diboson process (\ref{procWZ}),
employing one of the most recent ATLAS measurements
\cite{Aaboud:2017fgj}. Next, we turn to the dilepton production
process (\ref{proclept}), this process gives valuable
complementary information. Unlike the SSM, where there is
no $W$-$W'$ mixing, in the EGM we consider a non-zero
mixing $\xi$ in the analysis of
the $W'\to e\nu$ process. As described in Sec.~\ref{sect-cross},
this results in a modification of $\text{BR}(W'\to e\nu)$.

We compute the $W'$ production cross section at LO with PYTHIA~8.2
\cite{Sjostrand:2014zea}  at the LHC, $\sigma(pp\to W'X)$,
multiplied by the branching ratio into two leptons, $\ell\nu$
(here $\ell=e$), i.e., $\sigma(pp\to W'X) \times {\rm BR}(W'\to
e\nu)$, as a function of $M_{W'}$.  A mass-dependent $K$
factor is applied, based on NNLO QCD cross sections as calculated
with FEWZ~3.1 \cite{Gavin:2010az,Li:2012wna}. The $K$-factor varies
approximately from 1.3 to 1.1 for the range of $W'$ masses studied
in this analysis, namely from 0.5 to 6.0~TeV. The NNLO corrections
decrease with $W'$ boson masses up to around 4.5~TeV
\cite{Khachatryan:2016jww}. For higher $W'$ masses,
 the $K$-factor increases again and
becomes similar to the low-mass values.

The product of the NNLO $W'$  theoretical production cross section
and branching fraction, $\sigma(pp\to W'X) \times {\rm BR}(W'\to
e\nu)$, for the $W'$ boson for EGM strongly depends on the $W'$
mass, and is for illustrative purposes  given by thin solid curves,
in descending order corresponding to values of the mixing factor
$\xi$ from 0.0 to 0.01, as displayed in Fig.~\ref{fig5}. Further,
we compare the theoretical cross section $\sigma(pp\to W'X) \times {\rm
BR}(W'\to e\nu)$  with the upper limits of $\sigma_{95\%}\times
\text{BR}(W'\to e\nu)$ established by the ATLAS experiment
\cite{Aaboud:2017efa} for $36.1~\text{fb}^{-1}$. Qualitatively,
the decrease of the theoretical cross section with increasing
values of $\xi$ can be understood as follows: For increasing
$\xi$, the $W'\to WZ$ mode will at high mass $M_{W'}$ become more
dominant (as illustrated in Fig.~\ref{fig3}), and $\text{BR}(W'\to
e\nu)$ will decrease correspondingly.

Comparison of $\sigma(pp\to W'X) \times {\rm BR}(W'\to e\nu)$ vs
$\sigma_{95\%}\times \text{BR}(W'\to e\nu)$ displayed in
Fig.~\ref{fig5} allows us to read off an allowed mixing for a
given mass value, higher masses are allowed for smaller mixing,
for the reason stated above.  That comparison can be translated
into constraints on the two-dimensional $M_{W'}$-$\xi$ parameter
plane, as will be shown in the next
Sec.~\ref{sect:overall_constraints}.

The above  results are based on  data
corresponding to an integrated luminosity of 36.1 fb$^{-1}$ taken
by the ATLAS  collaboration at $\sqrt{s} = 13$ TeV in 2015 and
2016 \cite{Aaboud:2017efa}. However, recently the ATLAS collaboration
presented preliminary results on searching for a $W'$
boson conducted in the $W'\to \ell \nu$ channel (\ref{proclept})
based on 79.8 fb$^{-1}$ of $pp$ collision data collected
in 2015 (3.2 fb$^{-1}$), 2016 (33.0 fb$^{-1}$) and
2017 (43.6 fb$^{-1}$) at a centre-of-mass energy of $\sqrt{s} =
13$ TeV \cite{ATLAS:2018lcz}. While the latter analysis followed
closely the same procedure as in Ref.~\cite{Aaboud:2017efa}, the
sensitivity of the search presented in \cite{ATLAS:2018lcz} was
improved due to the inclusion of the 2017 dataset. Specifically,
this corresponds to an improvement of approximately $\sim$ 0.5 TeV in
mass reach compared to the previous ATLAS analysis
\cite{Aaboud:2017efa} which did not include the 2017
data and where a lower limit on the $W'_{\rm SSM}$ mass
of 5.2~TeV was set at 95\% CL. Those preliminary results
have not been included in
our present analysis.

\section{Summarizing constraints on the $W$-$W^\prime$ mixing}
\label{sect:overall_constraints}

\begin{figure}[!htb]
\begin{center}
\includegraphics[scale=0.75]{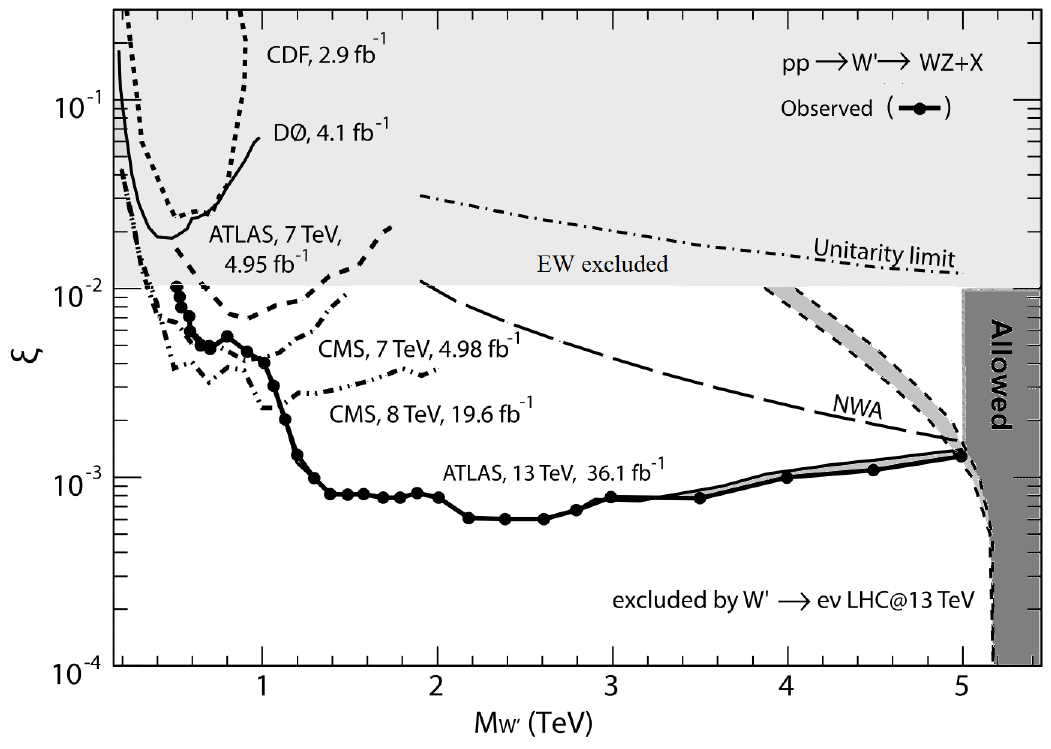}
\end{center}
\caption{95\%C.L. exclusion regions  in the
two-dimensional ($M_{W'}$, $\xi$) plane obtained after
incorporating indirect constraints from
electroweak precision data (the light shaded area at large $\xi$, labelled ``EW excluded''),
direct search constraints from the CDF and D0 collaborations
(Tevatron) in $p\bar{p}\to WZX$  as well as  from the LHC with  7,
8 and 13 TeV from diboson $W'\to WZ$ resonance search
data in $pp$ collisions. The region above each curve for the $WZ$ channel is excluded.
The steep curve labelled ``excluded by
$W'\to e\nu$ LHC@13 TeV'' shows the exclusion based on the
dilepton channel $pp\to e\nu X$. The light shaded
bands are defined like in Fig.~\ref{fig3}. The unitarity limit and the
upper bound for validity of the NWA are shown as dot-dashed and
long-dashed curves, respectively. The overall allowed region is
shown as a dark shaded area.} \label{fig6}
\end{figure}

As described above, both the diboson mode and the dilepton process
yield limits on the ($M_{W'}$, $\xi$) parameter space. These are
rather complementary, as shown in Fig.~\ref{fig6}, where we
collect these and other limits for the considered EGM model. The
limits arising from the diboson channel basically exclude
large values of $\xi$, strongest at intermediate masses
$M_{W'}\sim 2-4~\text{TeV}$. The limits arising from the dilepton
channel, on the other hand, basically exclude masses $M_{W'}\lsim
4-5.2~\text{TeV}$, with only a weak dependence on $\xi$. Also, we
show the unitarity limits discussed above, as well as the upper
bound for the validity of the NWA, indicated as dash-dotted and
long-dashed lines, respectively.

Interestingly, Fig.~\ref{fig6}, which is dedicated to the EGM
model, shows that at high $W'$ masses, the
limits on $\xi$ obtained from the ATLAS diboson resonance
production search at 13 TeV and at time-integrated luminosity of
 36.1 fb$^{-1}$ are substantially stronger than those derived from the
 low-energy electroweak data (EW), which are of order $\sim 10^{-2}$
\cite{Tanabashi:2018oca}.
For completeness, we display limits on the $W'$ parameters
from the CDF and D0 (Tevatron) as well as from ATLAS and CMS
obtained at 7  and 8 TeV of the LHC data taking in Run~I.
Fig.~\ref{fig6} shows that the experiments CDF and D0 at the Tevatron exclude EGM $W'$ boson with $\xi\gsim 2\cdot 10^{-2}$ and 0.6~TeV$<M_{W'}<$ 1 TeV   at the 95\% C.L., whereas LHC in Run~I improves those constraints excluding  $W'$ boson parameters at $\xi\gsim 2\cdot 10^{-3}$ in the mass range of 0.6~TeV$<M_{W'}<$ 2 TeV. \footnote{The latter ATLAS and CMS  constraints depicted in
Fig.~\ref{fig6} were obtained from resonant diboson production
data collected at low energies, 7 and 8 TeV, by translating
limits on the gauge coupling strength ${\cal C}$ vs $M_{W'}$
\cite{Aad:2013pdy,Chatrchyan:2012kk,Khachatryan:2014xja} onto
the $M_{W'}$-$\xi$ parameter plane, see Eq.~(\ref{Eq:define-xi}).}

In this note we combine the limits derived from $WZ$ production data
with those obtained from the $W'\to e\nu$ process in order to
significantly constrain the allowed region in the $M_{W'}$--$\xi$
parameter space to obtain the most stringent exclusion limits to date,
as illustrated in Fig.~\ref{fig6}. It is interesting to note that for
the EGM model under study the leptonic channel of (\ref{proclept})
provides the strongest bound on much of the
parameter space as shown in Fig.~\ref{fig6}, which is based on the
$e\nu$  and $WZ$ production data sets from the ATLAS experiment for
time-integrated luminosity of 36.1~fb$^{-1}$ at 13~TeV. In fact, the
parameter space for the $W'$ mass within the range of  $\sim
0.5-4$ TeV and $\xi<10^{-2}$ is basically excluded by  $W'\to e\nu$
process. In that mass range, $WZ$ production plays a complementary
role to the $W'\to e\nu$ process, excluding $W$-$W'$ mixing above $\xi\simeq 6\cdot 10^{-4}$. In addition, there is an area in parameter space within the $W'$ mass region ranging approximately from 4 TeV to 5 TeV and lying below the ``EW'' boundary  which is not excluded  solely by the process of Eq.~(\ref{proclept}). The exclusion of that area is achieved with  the process (\ref{procWZ}).

Furthermore, we could extrapolate the experimental sensitivity
curves for the $W'\to WZ$ decay channel for higher expected luminosity downwards by a factor of
$1/\sqrt D$, for the $M_{W'}$ mass range which was not statistically limited
(i.e., where there are events compatible with the SM background),
where $D$ is the ratio of the full integrated
luminosity of 139 fb$^{-1}$ that was collected by
the end of Run~II  \cite{ATLAS:2019vcr,ATLAS:2019cpi}, to the already analyzed integrated luminosity
of 36.1~fb$^{-1}$ in the ATLAS experiment.
It is clear that
further improvement on the constraining of this mixing can be
achieved from the analysis of such data.
At fixed $M_{W'}$ the exclusion constraint on $\xi$ scales as
$\sim\Lumint^{-1/4}$ when statistical errors dominate.
The increase of the time-integrated luminosity from 36.1~fb$^{-1}$ to
139 fb$^{-1}$ will allow to set stronger constraints on the mixing
angle $\xi$ by a factor of $\simeq
 1.4$.\footnote{Such scaling
law is also adopted for evaluation of the
$Z$-$Z'$ mixing strength vs $\Lumint$ in the process $pp\to Z' X\to
W^+W^- X$ \cite{Osland:2017ema,Bobovnikov:2018fwt}.}
We find that the LHC limits obtained at 13 TeV and time-integrated luminosity, $\Lumint=36.1$ fb$^{-1}$,
already improves on
the EW limits approximately by one order of magnitude.

Further improvement in placing limits on the $W'$ mass and $W$-$W'$ mixing parameter is feasible in fully-hadronic $WZ\to qqqq$ final states
using the full Run~II data set \cite{ATLAS:2019cpi}.  The recent analysis performed in \cite{ATLAS:2019cpi} is able to largely improve on the results
above for $\sim$ 36 fb$^{-1}$, mostly due to the use of novel reconstruction and analysis techniques.
Our fast and approximate estimation based on the full (preliminary) Run~II data set shows that the improvement in expected upper limits on $\xi$  for the $WZ$
channel at $\sim$ 3 TeV  is about a factor of two smaller than that presented here.
 However, a detailed analysis of the $W$-$W'$ mixing effects in the fully-hadronic $WZ\to qqqq$ final states is beyond the scope of the present paper and will be presented elsewhere.

\section{Concluding remarks}
\label{sect:conclusions}

Exploration of the the diboson $WZ$ and dilepton $e\nu$  production at
the LHC with the 13~TeV
data set allows to place stringent constraints on the $W$-$W'$ mixing
angle and $W'$ mass, $M_{W'}$. We derived such limits by
using data recorded
by the ATLAS detector at the CERN LHC, with integrated luminosity
of $\sim$ 36.1~fb$^{-1}$. By comparing the experimental limits to
the theoretical predictions for the total cross section of $W'$
resonant production and its subsequent decay into $WZ$ and $e\nu$ pairs, we
show that the derived constraints on the $W$-$W'$ mixing angle for
the EGM model are substantially improved
with respect to those obtained from the global analysis of low
energy electroweak data, as well as from the diboson production study
performed at the Tevatron and those based on the LHC Run~I.
Further constraining of this mixing can be achieved from the analysis of data
already collected by the end of Run~II, still to be analyzed.
The new LHC bound starts to become highly competitive with the
constraints coming from low-energy EW studies and after the LHC Run~I.

In addition, our work shows that accounting for the
contribution of the $W'$ boson decay channel,  $W'\to WH$, to the
total width $\Gamma_{W'}$ does not dramatically affect the bounds
on the mixing parameter $\xi$ obtained in the  scenario of
a vanishing $WH$ mode. Namely, it turns out that for
the higher $W'$ masses the constraints on
$W$-$W'$ mixing are relaxed by a few
relative percent for the $WZ$ channel as illustrated in Fig.~\ref{fig6}.
One should also note that in this paper, for the sake of compactness
of the graphic material, we limited ourselves to an analysis of
experimental data from the ATLAS detector only. Our further
analysis shows that the corresponding CMS data
\cite{Khachatryan:2016jww,Sirunyan:2017acf,Sirunyan:2018hsl} yields bounds on the mixing parameter
$\xi$ and the $W'$ boson mass that agree with the results based on
the ATLAS data.

\vspace*{0mm}

\section*{Acknowledgements}
We would like to thank V.~A. Bednyakov for helpful discussions.
This research has been partially supported by the Abdus Salam ICTP
(TRIL Programme). The work of PO has been supported by the
Research Council of Norway.


\raggedright

\end{document}